\documentclass[preprint]{sigplanconf}
\usepackage[utf8]{inputenc}
\usepackage[T1]{fontenc}
\usepackage{times}
\usepackage{qcourier}
 
\usepackage{graphicx}
\usepackage{hyperref}

\usepackage{relsize}
\usepackage{wasysym}
\usepackage{textcomp}
\usepackage{framed}
\usepackage[htt]{hyphenat}
\usepackage[usenames,dvipsnames]{color}
\hypersetup{bookmarks=true,bookmarksopen=true,bookmarksnumbered=true}
\IfFileExists{tocstyle.sty}{\usepackage{tocstyle}\usetocstyle{standard}}{}
\IfFileExists{CJK.sty}{\usepackage{CJK}}{}

\newcommand{\sectionNewpage}{}

\newcommand{\preDoc}{}
\newcommand{\postDoc}{}

\newcommand{\BookRef}[2]{\emph{#2}}
\newcommand{\ChapRef}[2]{\SecRef{#1}{#2}}
\newcommand{\SecRef}[2]{section~#1}
\newcommand{\PartRef}[2]{part~#1}
\newcommand{\BookRefUC}[2]{\BookRef{#1}{#2}}
\newcommand{\ChapRefUC}[2]{\SecRefUC{#1}{#2}}
\newcommand{\SecRefUC}[2]{Section~#1}
\newcommand{\PartRefUC}[2]{Part~#1}

\newcommand{\BookRefLocal}[3]{\hyperref[#1]{\BookRef{#2}{#3}}}
\newcommand{\ChapRefLocal}[3]{\hyperref[#1]{\ChapRef{#2}{#3}}}
\newcommand{\SecRefLocal}[3]{\hyperref[#1]{\SecRef{#2}{#3}}}
\newcommand{\PartRefLocal}[3]{\hyperref[#1]{\PartRef{#2}{#3}}}
\newcommand{\BookRefLocalUC}[3]{\hyperref[#1]{\BookRefUC{#2}{#3}}}
\newcommand{\ChapRefLocalUC}[3]{\hyperref[#1]{\ChapRefUC{#2}{#3}}}
\newcommand{\SecRefLocalUC}[3]{\hyperref[#1]{\SecRefUC{#2}{#3}}}
\newcommand{\PartRefLocalUC}[3]{\hyperref[#1]{\PartRefUC{#2}{#3}}}

\newcommand{\BookRefUN}[1]{\BookRef{}{#1}}

\newcommand{\BookRefLocalUN}[2]{\hyperref[#1]{\BookRefUN{#2}}}

\newcommand{\Scribtexttt}[1]{{\texttt{#1}}}

\newcommand{\planetName}[1]{PLane\hspace{-0.1ex}T}

\makeatletter
\def\empty@finalstrut#1{%
  \unskip\ifhmode\nobreak\fi\vrule\@width\z@\@height\z@\@depth\z@}
\def\no@strut{\global\setbox\@arstrutbox\hbox{%
    \vrule \@height\z@
           \@depth\z@
           \@width\z@}%
    \gdef\@endpbox{\empty@finalstrut\@arstrutbox\par\egroup\hfil}%
}%
\def\yes@strut{\global\setbox\@arstrutbox\hbox{%
    \vrule \@height\arraystretch \ht\strutbox
           \@depth\arraystretch \dp\strutbox
           \@width\z@}%
    \gdef\@endpbox{\@finalstrut\@arstrutbox\par\egroup\hfil}%
}%
\def\@mkpream#1{\@firstamptrue\@lastchclass6
  \let\@preamble\@empty\def\empty@preamble{\add@ins}%
  \let\protect\@unexpandable@protect
  \let\@sharp\relax\let\add@ins\relax
  \let\@startpbox\relax\let\@endpbox\relax
  \@expast{#1}%
  \expandafter\@tfor \expandafter
    \@nextchar \expandafter:\expandafter=\reserved@a\do
       {\@testpach\@nextchar
    \ifcase \@chclass \@classz \or \@classi \or \@classii \or \@classiii
      \or \@classiv \or\@classv \fi\@lastchclass\@chclass}%
  \ifcase \@lastchclass \@acol
      \or \or \@preamerr \@ne\or \@preamerr \tw@\or \or \@acol \fi}
\def\@addamp{%
  \if@firstamp
    \@firstampfalse
    \edef\empty@preamble{\add@ins}%
  \else
    \edef\@preamble{\@preamble &}%
    \edef\empty@preamble{\expandafter\noexpand\empty@preamble &\add@ins}%
  \fi}
\newif\iftw@hlines \tw@hlinesfalse
\def\@xhline{\ifx\reserved@a\hline
               \tw@hlinestrue
             \else\ifx\reserved@a\Hline
               \tw@hlinestrue
             \else
               \tw@hlinesfalse
             \fi\fi
      \iftw@hlines
        \aftergroup\do@after
      \fi
      \ifnum0=`{\fi}%
}
\def\do@after{\emptyrow[\the\doublerulesep]}
\def\emptyrow{\noalign\bgroup\@ifnextchar[\@emptyrow{\@emptyrow[\z@]}}
\def\@emptyrow[#1]{\no@strut\gdef\add@ins{\vrule \@height\z@ \@depth#1 \@width\z@}\egroup%
\empty@preamble\\
\noalign{\yes@strut\gdef\add@ins{\vrule \@height\z@ \@depth\z@ \@width\z@}}%
}
\def\tabrow#1{\noalign\bgroup\@ifnextchar[{\@tabrow{#1}}{\@tabrow{#1}[]}}
\def\@tabrow#1[#2]{\no@strut\egroup#1\ifx.#2.\\\else\\[#2]\fi\noalign{\yes@strut}}
\def\endpltstabular{\crcr\egroup\egroup \egroup}
\expandafter \let \csname endpltstabular*\endcsname = \endpltstabular
\def\pltstabular{\let\@halignto\@empty\@pltstabular}
\@namedef{pltstabular*}#1{\def\@halignto{to#1}\@pltstabular}
\def\@pltstabular{\leavevmode \bgroup \let\@acol\@tabacol
   \let\@classz\@tabclassz
   \let\@classiv\@tabclassiv \let\\\@tabularcr\@stabarray}
\def\@stabarray{\m@th\@ifnextchar[\@sarray{\@sarray[c]}}
\def\@sarray[#1]#2{%
  \bgroup
  \setbox\@arstrutbox\hbox{%
    \vrule \@height\arraystretch\ht\strutbox
           \@depth\arraystretch \dp\strutbox
           \@width\z@}%
  \@mkpream{#2}%
  \edef\@preamble{%
    \ialign \noexpand\@halignto
      \bgroup \@arstrut \@preamble \tabskip\z@skip \cr}%
  \let\@startpbox\@@startpbox \let\@endpbox\@@endpbox
  \let\tabularnewline\\%
    \let\@sharp##%
    \set@typeset@protect
    \lineskip\z@skip\baselineskip\z@skip
    \@preamble}

\makeatother

\newlength{\stabLeft}

\newenvironment{SingleColumn}{\begin{list}{}{\topsep=0pt\partopsep=0pt%
\listparindent=0pt\itemindent=0pt\labelwidth=0pt\leftmargin=0pt\rightmargin=0pt%
\itemsep=0pt\parsep=0pt}\item}{\end{list}}

\newcommand{\SCodePreSkip}{\vskip\abovedisplayskip}
\newcommand{\SCodePostSkip}{\vskip\belowdisplayskip}

\newcommand{\SVInsetPreSkip}{\vskip\abovedisplayskip}
\newcommand{\SVInsetPostSkip}{\vskip\belowdisplayskip}

\newcommand{\titleAndVersionAndAuthors}[3]{\title{#1\\{\normalsize \SVersionBefore{}#2}}\author{#3}\maketitle}

\newcommand{\titleAndEmptyVersionAndAuthors}[3]{\title{#1}\author{#3}\maketitle}

\newcommand{\SAuthor}[1]{#1}
\newcommand{\SAuthorSep}[1]{\qquad}
\newcommand{\SVersionBefore}[1]{Version }

\newcommand{\SNumberOfAuthors}[1]{}

\let\SOriginalthesubsection\thesubsection
\let\SOriginalthesubsubsection\thesubsubsection

\newcommand{\Ssection}[2]{\section[#1]{#2}\let\thesubsection\SOriginalthesubsection}
\newcommand{\Ssubsection}[2]{\subsection[#1]{#2}\let\thesubsubsection\SOriginalthesubsubsection}

\newcommand{\Ssectionstar}[1]{\section*{#1}\renewcommand*\thesubsection{\arabic{subsection}}\setcounter{subsection}{0}}

\newcommand{\Ssectionstarx}[2]{\Ssectionstar{#2}\addcontentsline{toc}{section}{#1}}

\newcounter{GrouperTemp}

\newcommand{\Snolinkurl}[1]{\nolinkurl{#1}}

\newcommand{\SAuthorinfo}[3]{#1}
\newcommand{\SAuthorPlace}[1]{#1}
\newcommand{\SAuthorEmail}[1]{#1}

\newcommand{\SConferenceInfo}[2]{}
\newcommand{\SCopyrightYear}[1]{}
\newcommand{\SCopyrightData}[1]{}
\newcommand{\Sdoi}[1]{}

\newcommand{\SCategory}[3]{}
\newcommand{\SCategoryPlus}[4]{}
\newcommand{\STerms}[1]{}
\newcommand{\SKeywords}[1]{}

\newcommand{\SSubtitle}[1]{{\bf #1}}

\newcommand{\NoteBox}[1]{\footnote{#1}}
\newcommand{\NoteContent}[1]{#1}

\newcommand{\FootnoteRef}[1]{}
\newcommand{\FootnoteTarget}[1]{}

\newcommand{\FootnoteBlockContent}[1]{}
\usepackage{ccaption}

\makeatletter
\@ifundefined{belowcaptionskip}{}{}
\makeatother

\newcommand{\Legend}[1]{~

                        \hrule width \hsize height .33pt
                        \vspace{4pt}
                        \legend{#1}}

\newcommand{\FigureTarget}[2]{#1}
\newcommand{\FigureRef}[2]{\hyperref[#2]{#1}}

\newlength{\FigOrigskip}
\FigOrigskip=\parskip

\newcommand{\FigureSetRef}{\refstepcounter{figure}}

\newenvironment{Figure}{\begin{figure}\FigureSetRef}{\end{figure}}
\newenvironment{FigureMulti}{\begin{figure*}[t!p]\FigureSetRef}{\end{figure*}}

\newenvironment{Centerfigure}{\begin{Xfigure}\centering\item}{\end{Xfigure}}

\newenvironment{Xfigure}{\begin{list}{}{\leftmargin=0pt\topsep=0pt\parsep=\FigOrigskip\partopsep=0pt}}{\end{list}}

\newenvironment{FigureInside}{}{}

\newcommand{\Centertext}[1]{\begin{center}#1\end{center}}

\newenvironment{AutoBibliography}{\begin{small}}{\end{small}}
\newcommand{\Autobibentry}[1]{\hspace{0.05\linewidth}\parbox[t]{0.95\linewidth}{\parindent=-0.05\linewidth#1\vspace{1.0ex}}}

\usepackage{calc}
\newlength{\ABcollength}

\def\SXtitle#1{\title{\let\SSubtitle\SSubtitleDrop#1}\SExtractSubtitle#1\SExtractSubtitleDone}
\def\SSubtitleDrop#1{}
\def\SExtractSubtitleDone {}
\def\SExtractSubtitle{\futurelet\next\SExtractSubtitleX}
\def\SExtractSubtitleX#1{\ifx#1\SSubtitle \let\Snext\SWithSubtitle \else \let\Snext\SExtractSubtitleY \fi \Snext}
\def\SExtractSubtitleY{\ifx\next\SExtractSubtitleDone \let\Snext\relax \else \let\Snext\SExtractSubtitle \fi \Snext}
\def\SWithSubtitle#1{\subtitle{#1}\SExtractSubtitle}

\renewcommand{\titleAndVersionAndAuthors}[3]{\SXtitle{#1}#3\maketitle}
\renewcommand{\titleAndEmptyVersionAndAuthors}[3]{\titleAndVersionAndAuthors{#1}{#2}{#3}}

\def\SAuthor#1{\SAutoAuthor#1\SAutoAuthorDone{#1}}
\def\SAutoAuthorDone#1{}
\def\SAutoAuthor{\futurelet\next\SAutoAuthorX}
\def\SAutoAuthorX{\ifx\next\SAuthorinfo \let\Snext\relax \else \let\Snext\SToAuthorDone \fi \Snext}
\def\SToAuthorDone{\futurelet\next\SToAuthorDoneX}
\def\SToAuthorDoneX#1{\ifx\next\SAutoAuthorDone \let\Snext\SAddAuthorInfo \else \let\Snext\SToAuthorDone \fi \Snext}
\newcommand{\SAddAuthorInfo}[1]{\authorinfo{#1}{}{}}

\renewcommand{\SAuthorinfo}[3]{\authorinfo{#1}{#2}{#3}}
\renewcommand{\SAuthorSep}[1]{}

\renewcommand{\SConferenceInfo}[2]{\conferenceinfo{#1}{#2}}
\renewcommand{\SCopyrightYear}[1]{\copyrightyear{#1}}
\renewcommand{\SCopyrightData}[1]{\copyrightdata{#1}}
\renewcommand{\Sdoi}[1]{\doi{#1}}

\renewcommand{\SCategory}[3]{\category{#1}{#2}{#3}}
\renewcommand{\SCategoryPlus}[4]{\category{#1}{#2}{#3}[#4]}
\renewcommand{\STerms}[1]{\terms{#1}}
\renewcommand{\SKeywords}[1]{\keywords{#1}}

\doi{}
\begin{document}
\preDoc
\titleAndEmptyVersionAndAuthors{Generating 56{-}bit passwords using Markov Models (and Charles Dickens)}{}{\SNumberOfAuthors{1}\SAuthor{\SAuthorinfo{John Clements}{\SAuthorPlace{Cal Poly San Luis Obispo}}{\SAuthorEmail{clements@brinckerhoff.org}}}}
\label{t:x28part_x22Generatingx5f56x2dbitx5fpasswordsx5fusingx5fMarkovx5fModelsx5fx5fandx5fCharlesx5fDickensx5fx22x29}

\begin{abstract}We describe a password generation scheme based on Markov models
built from English text (specifically, Charles Dickens{'} \textit{A Tale
Of Two Cities}). We show a (linear{-}running{-}time) bijection between
 random bitstrings
of any desired length and generated text, ensuring that all passwords
are generated with equal probability. We observe that the generated
passwords appear to strike a reasonable balance between memorability
and security. Using the system, we get 56{-}bit passwords like
\Scribtexttt{The cusay is wither{\hbox{\texttt{?}}}" t}, rather than passwords like \Scribtexttt{tQ\$\%Xc4Ef}.\end{abstract}

\sectionNewpage

\Ssection{Introduction}{Introduction}\label{t:x28part_x22Introductionx22x29}

Users are very bad at choosing passwords.

In order to precisely quantify just how bad they are (and
how much better we would like to be), we use the standard
measure of "bits of entropy", due to Shannon~(Shannon 1948). As an example,
a password chosen randomly from a set of 1024 available
passwords would exhibit 10 bits of entropy, and more
generally, one chosen at random from a set of size \relax{\(S\)}
will exhibit \relax{\(log_2{S}\)} bits of entropy.

In a system using user{-}chosen passwords, some passwords will be chosen
more frequently than others. This makes it much harder to characterize
the {``}average{''} entropy of the passwords, but analysis by Bonneau
of more than 65 million Yahoo passwords suggests that an attacker
that is content to crack 25\% of passwords can do so by trying a pool
whose size is 25\% of \relax{\(2^{17.6}\)}. That is, the least secure quarter of
users are as safe as they would be with a randomly generated password
with 17.6 bits of entropy~(Bonneau 2012).

To see just how terrible this is, observe that we can easily construct
a pool of 77 password{-}safe characters\NoteBox{\NoteContent{viz: abcdefghijklmnopqrstuvwxyz
ABCDEFGHIJKLMNOPQRSTUVWXYZ
1234567890!{\char'136}{-}=+[]@\#\$\%\&*()}},
so that a randomly generated password containing \relax{\(n\)} characters will
contain \relax{\(n\log_2(77)\)} or approximately \relax{\(6.25n\)} bits of entropy,
and that the aforementioned 50\% of users would be better served by a
password of three randomly generated characters. To better gauge this
difficulty, observe this set of 8 randomly generated e{-}character passwords:\NoteBox{\NoteContent{Throughout this paper, in the spirit of even{-}handedness and honesty, we have
been careful to run each example only once, to avoid the tendency to {``}cherry{-}pick{''}
examples that suit our points.}}

\begin{SingleColumn}\Scribtexttt{tBJ}

\Scribtexttt{fZX}

\Scribtexttt{evA}

\Scribtexttt{8Fy}

\Scribtexttt{MHr}

\Scribtexttt{=qe}

\Scribtexttt{f]w}

\Scribtexttt{YxU}\end{SingleColumn}

We conjecture that most users could readily memorize one of these.\NoteBox{\NoteContent{Please
don{'}t use these passwords, or any other password printed in this paper. These
passwords are officially toast.}}

Unfortunately, we need to set the target substantially higher. One
standard attack model assumes that attackers will have access to
encrypted passwords for offline testing, but that the password
encryption scheme will use {``}key stretching,{''} a method of relying
on expensive{-}to{-}compute hashes in order to make checking passwords{---}and
therefore, guessing passwords{---}more expensive.

Bonneau and Schechter suggest that under these constraints,
and the further assumption that key{-}stretching can be increased to compensate
for ever{-}faster machines, a password with 56 bits of entropy might well
be considered adequate for some time to come~(Bonneau and Schechter 2014).

The most straightforward way to achieve this goal is with randomly generated
passwords. That is, users are assigned passwords by the system, rather
than being allowed to choose their own. In fact, this was standard practice
until approximately 1990~(Adams et al. 1997), when user convenience was
seen to to trump security.

Today, the general assumption{---}evidenced by the lack of systems using randomly
assigned passwords{---}is that users cannot be expected to recall secure passwords.
Bonneau and Schechter~(Bonneau and Schechter 2014) challenge this, and describe a study in which users
were recruited for an experiment in which they were unwittingly learning
to type a 56{-}bit password.\NoteBox{\NoteContent{Later interviews suggested that
some of them might have deduced the experiment{'}s true goal.}} This experiment
used \textit{spaced repetition}~(Cepeda et al. 2006; Ebbinghaus 1885), and found that users learned their passwords after
a median of 36 logins, and that three days later, 88\% recalled their passwords
precisely, although 21\% admitted having written them down.

\sectionNewpage

\Ssection{How to Randomly Generate Passwords?}{How to Randomly Generate Passwords?}\label{t:x28part_x22Howx5ftox5fRandomlyx5fGeneratex5fPasswordsx5fx22x29}

If we{'}re convinced that random passwords are a good idea, and that recalling
a 56{-}bit password is at least within the realm of possibility, we must try
to find a set of passwords (more specifically, a set of \relax{\(2^{56}\)} passwords)
that are as memorable as possible.

We should acknowledge at the outset that there are many password schemes that
use passwords that are not simply alphanumeric sequences. We acknowledge the
work that{'}s gone into these approaches, and we regard these schemes as
outside the scope of this paper.

\Ssubsection{Random Characters}{Random Characters}\label{t:x28part_x22Randomx5fCharactersx22x29}

The first and most natural system is to generate passwords by choosing random
sequences of characters from a given set, as described before.  In order to
see what a 56{-}bit password might look like in such a system, consider the
following set of eight such passwords:

\begin{SingleColumn}\Scribtexttt{Ocd{\hbox{\texttt{!}}}SG3aU}

\Scribtexttt{)u)4OlXt\%}

\Scribtexttt{tQ\$\%Xc4Ef}

\Scribtexttt{TH9H*kt7{\char'136}}

\Scribtexttt{@f7naKFpx}

\Scribtexttt{K+UKdf{\char'136}7c}

\Scribtexttt{S{\char'136}UhiU\#cm}

\Scribtexttt{usCGQZ)p{-}}\end{SingleColumn}

In this system, a single randomly generated password has an entropy of 56.4 bits.

Naturally, a different alphabet can be used, and this will affect memorability. For
instance, we use an alphabet containing only one and zero:

\begin{SingleColumn}\Scribtexttt{11011111100111010101111100111010100010000110000011110110}

\Scribtexttt{10010110011110100010000011001111111000101100110010001001}

\Scribtexttt{11101101110001000001011001011110000111000101100001011101}

\Scribtexttt{11101001000011010100110010011000111000110011110001010011}

\Scribtexttt{00110110011001110011000111001111111011011101010010111000}

\Scribtexttt{11001001101011110111101010100100010001110111111111111101}

\Scribtexttt{10000101110111100101010111010000111110111010110111100100}

\Scribtexttt{11101010100010011010000101010000101010110010110110001001}\end{SingleColumn}

In this system, each password is 56 characters long, and has exactly 56 bits of
entropy. We conjecture that passwords such as these would be difficult to memorize.

\Ssubsection{Random Words}{Random Words}\label{t:x28part_x22Randomx5fWordsx22x29}

Alternatively, many more than six bits can be encoded in each character, if we take
as elements of our alphabet not single letters but rather words, or syllables.

The first of these, perhaps best known through the "Horse Battery Staple" XKCD comic
~(Monroe 2011), suggests that we use a word list, and choose from a
small set of word separators to obtain a bit of extra entropy. Using the freely
available RIDYHEW word list~(Street ???), we can obtain 18.8 bits of entropy for
each word, plus 2 bits for each separator. In order to reach the 56{-}bit threshold,
we must therefore use three of each, for a total of 62 bits of entropy. Here are
eight examples:

\begin{SingleColumn}\Scribtexttt{reelman,phymas{-}quelea;}

\Scribtexttt{leapful;bubinga;morsures{-}}

\Scribtexttt{orientalised;liging{-}isographs{-}}

\Scribtexttt{molecule{-}charcoallier{-}foxings,}

\Scribtexttt{plaquette{\hbox{\texttt{.}}}cultivates{\hbox{\texttt{.}}}agraphobia{-}}

\Scribtexttt{mewsed;gasmasking;pech;}

\Scribtexttt{metencephalic{\hbox{\texttt{.}}}gulf{\hbox{\texttt{.}}}layoff;}

\Scribtexttt{kinematicses{-}pyknosomes;delineate{\hbox{\texttt{.}}}}\end{SingleColumn}

Our observation (at the time of the comic{'}s release) was that these sequences
did not seem to be substantially nicer than the simple alphanumeric sequences,
due in large part to the use of words like {``}pyknosomes,{''} {``}quelea,{''} and {``}phymas.{''}

\Ssubsection{Random Syllables}{Random Syllables}\label{t:x28part_x22Randomx5fSyllablesx22x29}

A number of other schemes have attempted to split the difference between random
characters and random words by using random syllables. One such scheme was adopted
by the NIST~(NIST 1993), although it was later found to be broken, in that
it generated passwords with different probabilities~(Ganesan and Davies 1994). Despite
this, it is not difficult to devise a scheme in which all syllables are equally likely
to be generated.

One example of such a scheme is given by Leonhard and Venkatakrishnan~(Leonhard and Venkatakrishnan 2007).
They generate words by choosing from a set of 13 templates, where each template indicates
which characters must be consonants, and which characters must be vowels. So, for instance,
one of the templates is "abbabbaa", indicating that the first character must be a vowel,
the second two must be consonants, and so forth. Each consonant is chosen from a fixed set,
as is each vowel. The resulting words have 30.8 bits of entropy; in order to achieve the
needed 56, we can simply choose two of them.

Here are eight such examples:

\begin{SingleColumn}\Scribtexttt{kuyivavo rastgekoe}

\Scribtexttt{phoymasui nupiirji}

\Scribtexttt{ifstaezfa ihleophi}

\Scribtexttt{stifuyistu apibzaco}

\Scribtexttt{iholeyza gohwoopha}

\Scribtexttt{ebyexloi stustoijsto}

\Scribtexttt{maiwixdi enjujvia}

\Scribtexttt{dophaordu ostchichbou}\end{SingleColumn}

\sectionNewpage

\Ssection{Driving Nonuniform Choice using Bit Sources}{Driving Nonuniform Choice using Bit Sources}\label{t:x28part_x22Drivingx5fNonuniformx5fChoicex5fusingx5fBitx5fSourcesx22x29}

One characteristic of all of the approaches seen thus far is that they guarantee
that every password is chosen with equal probability, using a simple approach.
Specifically, password generation proceeds by making a fixed number of choices
from a fixed number of a fixed set of elements.

Specifically, the first scheme generates a password by making exactly ten choices
from sets of size 77, for all passwords. The last scheme is also careful to ensure
the same number of vowels and consonants in each template, meaning that password
generation always involves one choice from a set of size 13 followed by four choices
from a set of size 5 (the vowels) and four choices from a set of size 22, followed
by a second round of each of these (in order to generate a second word).  For all of
these schemes, every possible word is generated with equivalent probability. This
property is crucial, since a system that generates some passwords with higher probability{---}such
as the scheme adopted by the NIST~(NIST 1993){---}means that by focusing on
more probable passwords, attackers can gain leverage.

This approach has a cost, though. In such a scheme, it is not possible to {``}favor{''} certain
better{-}sounding or more{-}memorable passwords by biasing the system toward their selection; such
a bias would increase the probability of certain passwords being generated, and thereby
compromise the system.

\Ssubsection{Another Way}{Another Way}\label{t:x28part_x22Anotherx5fWayx22x29}

However, there is another way of guaranteeing that each password is generated with equal likelihood.
If we can establish a (computable) bijection between the natural numbers in the range \relax{\([0, \ldots ,N)\)}
and a set of passwords, then we can easily guarantee that each password is generated with
equal probability by directly generating a random natural number, and then mapping it to
the corresponding password.

In order to make such a scheme work, we must show that mapping is indeed a bijection,
implying that no two numbers map to the same password.

\Ssubsection{Using Bits to Drive a Model}{Using Bits to Drive a Model}\label{t:x28part_x22Usingx5fBitsx5ftox5fDrivex5fax5fModelx22x29}

This idea opens up a new way to generate passwords. Rather than making a sequence of independent
choices, we can build a model that draws randomness from a given sequence of bits. That is,
we first generate a sequence of 56 random bits, and then use this as a stream of randomness
to determine the behavior of a pseudo{-}random algorithm.  If the stream of bits represents the
only source of (apparent) nondeterminism, then in fact the algorithm is deterministic, and
indeed determined entirely by the given sequence of bits.

Using this approach, we can lift the restriction (all choices must be equally likely) that
has dogged the creation of memorable or idiomatic{-}sounding password generators.

Specifically, our chosen non{-}uniform approach uses a Markov model, built from Charles Dickens{'}
\textit{A Tale of Two Cities.} We conjecture that this choice is not a critical one.

\sectionNewpage

\Ssection{Markov Models}{Markov Models}\label{t:x28part_x22Markovx5fModelsx22x29}

In its simplest form, a Markov modelis simply a nondeterministic state machine.
The model contains a set of states, and a set of transitions. Each transition has a probability
associated with it, and we have the standard invariant that the sum of the probabilities of
the transitions from the given states sum to one.

For our work, we built markov models from the sequences of characters\NoteBox{\NoteContent{when we say
characters, we mean letters in the alphabet, not the fictional subjects of the novel...}} in Charles Dickens{'}
A Tale of Two Cities~(Dickens 1859). One choice that we faced was
how many characters to include in each state. For the sake of the following examples, we will
fix this number at two.

To build the model, then, consider every pair of adjacent characters in the book. For instance,
\Scribtexttt{"ca"} is one such pair of characters. Then, consider every character that follows this
pair, and count how many times each occurs. This generates the distribution shown in
figure~\FigureRef{1}{t:x28counter_x28x22figurex22_x22cax2dhistogramx22x29x29}:

\begin{Figure}\begin{Centerfigure}\begin{FigureInside}\raisebox{-0.19999999999998863bp}{\makebox[160.0bp][l]{\includegraphics[trim=2.4000000000000004 2.4000000000000004 2.4000000000000004 2.4000000000000004]{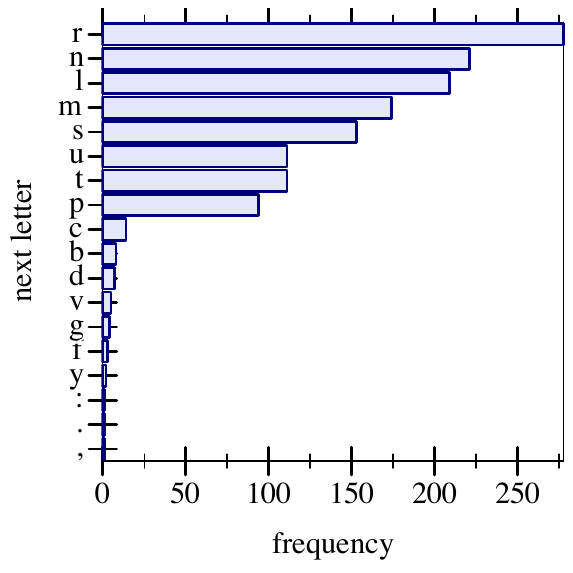}}}\end{FigureInside}\end{Centerfigure}

\Centertext{\Legend{\FigureTarget{\label{t:x28counter_x28x22figurex22_x22cax2dhistogramx22x29x29}Figure~1: }{t:x28counter_x28x22figurex22_x22cax2dhistogramx22x29x29}distribution of letters following\Scribtexttt{"ca"}}}\end{Figure}

In order to generate idiomatic text from this model, then, we should observe these
distributions. That is, if the last two characters were \Scribtexttt{"ca"}, the next character
should be an \Scribtexttt{"r"} with probability 278/1397.

How should we make this choice? One way would be to draw enough bits (11) from our pool
to get a number larger
than 1397, and then, say, pick the letter \Scribtexttt{"r"} if the number is less than 278. Note, though,
that while our program will be deterministic (since it gets its randomness from the
stream of given bits), it will \textit{not} represent a bijection, since (at least) 278 of
the 2048 possible choices all go to the same state.

To solve this, we need a way of drawing fewer bits to make more common choices, and drawing
more bits to make rarer ones.

Fortunately, this is exactly the problem that Huffman trees solve!

\sectionNewpage

\Ssection{Huffman Trees}{Huffman Trees}\label{t:x28part_x22Huffmanx5fTreesx22x29}

Huffman trees~(Huffman and others 1952) are generally used in compression. The basic idea is that
we can build a binary tree where more{-}common choices are close to the root, and less{-}common
choices are further from the root.

The standard construction algorithm for Huffman trees proceeds by coalescing; starting with
a set of leaves with weights, we join together the two least{-}weighty leaves into a branch
whose weight is the sum of its children. We then continue, until at last we{'}re left with
just one tree.

As an example, we can consider the distribution given above. In this case, there are several
characters (the comma, the period, and the colon) that occur just once.  We would therefore
combine two of these (the comma and the period, say) into a branch with weight two and two
children, the comma and period leaves. Next, we would combine the colon (the only tree left
with weight one) with either the \Scribtexttt{"y"} or the branch formed in the previous step; each
has weight two. The result would have weight three.

Proceeding in this way, we arrive at the tree shown in figure~\FigureRef{2}{t:x28counter_x28x22figurex22_x22nextx2dletterx2dtreex22x29x29}.

\begin{Figure}\begin{Centerfigure}\begin{FigureInside}\raisebox{-0.6499999999999773bp}{\makebox[156.0bp][l]{\includegraphics[trim=2.4000000000000004 2.4000000000000004 2.4000000000000004 2.4000000000000004]{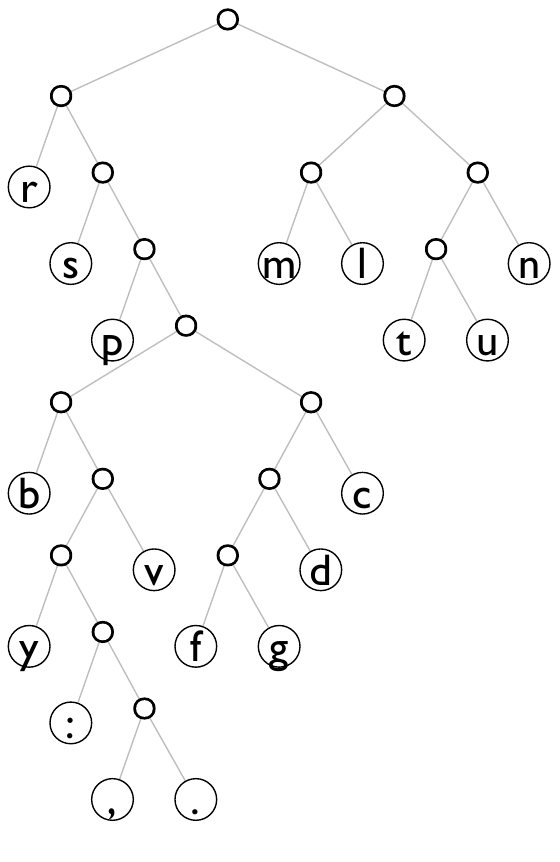}}}\end{FigureInside}\end{Centerfigure}

\Centertext{\Legend{\FigureTarget{\label{t:x28counter_x28x22figurex22_x22nextx2dletterx2dtreex22x29x29}Figure~2: }{t:x28counter_x28x22figurex22_x22nextx2dletterx2dtreex22x29x29}Huffman tree encoding next{-}letter choice from state \Scribtexttt{"ca"}}}\end{Figure}

If this tree were to be used in compression, we would
represent the transition to the letter \Scribtexttt{"r"} using two
bits, a zero and a zero (if we use zeros to denote left
branches). The transition to the next most likely letter,
\Scribtexttt{"l"}, would be represented as one{-}zero{-}one. Note that
less common choices are encoded using larger numbers of
bits.

We are not interested in compression, but in generation. For
this use case, we imagine that we are "decoding" the random
bit stream. So, for instance, if the random bit stream
contains the bits (0100110), we would use the first six bits
to reach the leaf \Scribtexttt{"c"}, and leave the remaining zero in
the stream.

Once we{'}ve reached a character, we may add this character to
the output stream. In order to continue, we must then start
again, in the new state.  If, for instance, the \Scribtexttt{"l"}
were chosen, we would now be in the state corresponding to
the letter pair \Scribtexttt{"al"}, and we would begin again.

Consider once more the problem of proving that this is a
bijection. In contrast to the earlier scheme, note that if
two bit streams differ first at (say) bit \relax{\(n\)}, then the
character that is output at that point in the model{'}s
operation is guaranteed to be different. This ensures that
each bit stream corresponds to a different output. To see
the other half of the bijection, we observe that given a
model{'}s output, we can simply run the "compression"
algorithm to obtain the sequence of bits that generated it.

\Ssubsection{Running Out of Bits}{Running Out of Bits}\label{t:x28part_x22Runningx5fOutx5fofx5fBitsx22x29}

One minor complication arises in that the given scheme is
not guaranteed to end "neatly". That is, the model may have
only partially traversed a huffman tree when the end of the
input bit stream is reached. We can easily solve this by
observing the bijection between streams of 56 randomly
generated bits and the infinite stream of bits whose first
56 bits are randomly generated and whose remaining bits are
all {``}zero{''}, in much the same way that an integer is not
changed by prepending an infinite stream of zeros. This
allows us to implement a bit generator that simply defaults
to {``}zero{''} when all bits are exhausted. In fact, the model
could continue generating text, but there{'}s no need to do
so, since the 56 random bits have already been used.

\Ssubsection{The Forbidden state}{The Forbidden state}\label{t:x28part_x22Thex5fForbiddenx5fstatex22x29}

Can our Markov model get stuck? This can occur if there is a
state with no outgoing transition. Fortunately, the
construction of the tree guarantees there will always be at
least one transition... except for the final characters
of the file. If this sequence occurs only once in the text
file, it{'}s conceivable that the model could get stuck. This
problem can easily be solved, though, by considering the source
document to be {``}circular,{''} and adding a transition from the
final state to the file{'}s initial character.

\Ssubsection{Choosing a Markov Model}{Choosing a Markov Model}\label{t:x28part_x22Choosingx5fax5fMarkovx5fModelx22x29}

In our examples thus far, we have chosen to use exactly two
characters as the states in the Markov model.  This is by no
means the only choice. We can easily use one character, or
three or four.

The tradeoff is fairly clear: using shorter
character{-}strings results in strings that sound less like
English, and using longer character{-}strings results in
strings that more like English. There is, however, a price;
the idiomaticity of the resulting strings results from a
lower "compression", measured in bits per character. That
is, the one{-}character markov model results in short strings,
and the three{-} and four{-}character models result in longer
ones. Naturally, all of the given models have the randomness
properties we{'}ve shown for the two{-}character ones, and users
may certainly choose a three{-} or four{-}character model, if
they find that the increase in memorability compensates for
the increase in length.

A final note concerns the selection of the initial state.
We{'}ve chosen simply to start with the appropriate{-}length
substring of "The ". Naturally, the starting state could be
chosen at random, to obtain slightly shorter strings.

\sectionNewpage

\Ssection{Examples}{Examples}\label{t:x28part_x22Examplesx22x29}

The proof is in the pudding! Let{'}s see some examples.

First, we generate strings using the one{-}character Markov model:

\begin{SingleColumn}\Scribtexttt{Tenon thempea co ts}

\Scribtexttt{Te od " perdy, wil}

\Scribtexttt{Thalivares youety}

\Scribtexttt{T{\hbox{\texttt{.}}}) reait dean,}

\Scribtexttt{Tr,{\textquotesingle}ser h Lof owey}

\Scribtexttt{Tempr{\hbox{\texttt{.}}}" gedolam,}

\Scribtexttt{Te cty se d y Mr,{-}}

\Scribtexttt{Tere th, Fand ry{\hbox{\texttt{.}}}"}\end{SingleColumn}

These may be seen to be short, but contain challenging sequences, such as \Scribtexttt{cty se d y}.

Next, strings generated using the two{-}character Markov model:

\begin{SingleColumn}\Scribtexttt{Therfur, unappen{\hbox{\texttt{.}}} So}

\Scribtexttt{Therying hant abree,}

\Scribtexttt{The cusay is wither{\hbox{\texttt{?}}}" t}

\Scribtexttt{The greed hispefters and}

\Scribtexttt{The as obe so yon ters}

\Scribtexttt{Thad gre strow; agamo}

\Scribtexttt{Thereakentin town ing{\hbox{\texttt{.}}}" "MO}

\Scribtexttt{Their, anyte{\hbox{\texttt{!}}}{\textquotesingle} hat," "te"}\end{SingleColumn}

These are slightly longer, but much more pronounceable, and appear substantially
more memorable.

Next, strings generated using the three{-}character Markov model:

\begin{SingleColumn}\Scribtexttt{Ther highly to a vice of eart}

\Scribtexttt{Then," suspeakings beers ways}

\Scribtexttt{They, anythis, int founged mad}

\Scribtexttt{They{\hbox{\texttt{?}}}" "If, who waite any," mul}

\Scribtexttt{The moritiour him; businenl}

\Scribtexttt{Thensuspellectiver fur}

\Scribtexttt{Then him do nown wilty," res}

\Scribtexttt{The fix, buse hand, followest{\hbox{\texttt{.}}}"}\end{SingleColumn}

These are far more English{-}like, with many actual words. As a side note, the
phrases generated here and in by the prior two{-}character model appear almost
archaic, with words like "waite," "nown," and "yon". Naturally, these are longer
than the prior set.

Finally, strings generated using the four{-}character Markov model:

\begin{SingleColumn}\Scribtexttt{The tile fareweloped, and ever p}

\Scribtexttt{The shing it nother to delve w}

\Scribtexttt{The found, Sydney Carton wreckles Evremonds{\hbox{\texttt{.}}}}

\Scribtexttt{}\mbox{\hphantom{\Scribtexttt{xx}}}\Scribtexttt{Su}

\Scribtexttt{The snorting ever in by turbed t}

\Scribtexttt{The receive a year{\hbox{\texttt{.}}}" To appeality a}

\Scribtexttt{The back understitch ther{\textquotesingle}s the}

\Scribtexttt{The wrong and, here{\hbox{\texttt{!}}}"{-}{-}Mr{\hbox{\texttt{.}}} Calm info}

\Scribtexttt{The diffidelicitizen aparticulous timo}\end{SingleColumn}

At this point, it{'}s fairly clear what the source is; Sydney Carton appears by name.
In addition, you get some fairly interesting neologisms{---}in this case, "diffidelicitizen."
It{'}s not a word, but maybe it should be. Also, we see some people that aren{'}t actually
in the book, including "Mr. Calm info."

\sectionNewpage

\Ssection{Choice of Corpus}{Choice of Corpus}\label{t:x28part_x22Choicex5fofx5fCorpusx22x29}

Naturally, the choice of \textit{A Tale of Two Cities} is largely arbitrary; any
corpus of reasonable length will suffice. One intriguing possibility would be
to choose the full text of all of the e{-}mails in a particular user{'}s history.
This text would presumably reflect the style of text that a particular user
is accustomed to read and write, and should in principle be extraordinarily
memorable.  Note that the security of the system is entirely independent of
the chosen corpus; our attack model assumes that the attacker already has
the full text of the corpus.

\sectionNewpage

\Ssection{Related Work}{Related Work}\label{t:x28part_x22Relatedx5fWorkx22x29}

There are many, many works that describe passwords. We have cited Bonneau{'}s
work before, and we will do so again here, as this work was enormously
informative~(Bonneau and Schechter 2014). We have also already described the work contained
in many other related projects~(Leonhard and Venkatakrishnan 2007; NIST 1993).

To our knowledge, however, there is no other work that uses a bit source to drive
huffman decoding to drive a markov model, thereby enabling generation of
pronounceable text without the (heretofore) attendant lack of equi{-}probability.

\sectionNewpage

\Ssection{Future Work}{Future Work}\label{t:x28part_x22Futurex5fWorkx22x29}

There{'}s a giant piece of future work here: specifically, we wave our hands
and suggest that our passwords are more memorable than those generated by
other schemes. Naturally, a claim like this cannot simply be taken as
true; we must conduct a test to verify this claim.

We are currently building the tools to allow us to conduct this study.

\sectionNewpage

\Ssectionstarx{Bibliography}{Bibliography}\label{t:x28part_x22docx2dbibliographyx22x29}

\begin{AutoBibliography}\begin{SingleColumn}\label{t:x28autobib_x22Anne_Adamsx2c_Martina_Angela_Sassex2c_and_Peter_LuntMaking_passwords_secure_and_usableIn_Procx2e_People_and_Computers_XIIx2c_ppx2e_1x2dx2d191997x22x29}\Autobibentry{Anne Adams, Martina Angela Sasse, and Peter Lunt. Making passwords secure and usable. In \textit{Proc. People and Computers XII}, pp. 1{--}19, 1997.}

\label{t:x28autobib_x22Joseph_BonneauThe_science_of_guessingx3a_analyzing_an_anonymized_corpus_of_70_million_passwordsIn_Procx2e_2012_IEEE_Symposium_on_Security_and_Privacy2012x22x29}\Autobibentry{Joseph Bonneau. The science of guessing: analyzing an anonymized corpus of 70 million passwords. In \textit{Proc. 2012 IEEE Symposium on Security and Privacy}, 2012.}

\label{t:x28autobib_x22Joseph_Bonneau_and_Stuart_SchechterTowards_reliable_storage_of_56x2dbit_secrets_in_human_memoryIn_Procx2e_Procx2e_USENIX_Security2014x22x29}\Autobibentry{Joseph Bonneau and Stuart Schechter. Towards reliable storage of 56{-}bit secrets in human memory. In \textit{Proc. Proc. USENIX Security}, 2014.}

\label{t:x28autobib_x22Nicholas_Jx2e_Cepedax2c_Harold_Pashlerx2c_Edward_Vulx2c_John_Tx2e_Wixtedx2c_and_Doug_RohrerDistributed_practice_in_verbal_recall_tasksx3a_A_review_and_quantitative_synthesisPsychological_Bulletin_132x283x292006x22x29}\Autobibentry{Nicholas J. Cepeda, Harold Pashler, Edward Vul, John T. Wixted, and Doug Rohrer. Distributed practice in verbal recall tasks: A review and quantitative synthesis. \textit{Psychological Bulletin} 132(3), 2006.}

\label{t:x28autobib_x22Charles_DickensA_Tale_of_Two_CitiesChapman_x26_Hall1859x22x29}\Autobibentry{Charles Dickens. A Tale of Two Cities. Chapman \& Hall, 1859.}

\label{t:x28autobib_x22Hermann_Ebbinghausxdcber_das_gedxe4chtnisx3a_untersuchungen_zur_experimentellen_psychologieDuncker_x26_Humblot1885x22x29}\Autobibentry{Hermann Ebbinghaus. \"{U}ber das ged\"{a}chtnis: untersuchungen zur experimentellen psychologie. Duncker \& Humblot, 1885.}

\label{t:x28autobib_x22Ravi_Ganesan_and_Chris_DaviesA_new_attack_on_random_pronounceable_password_generatorsIn_Procx2e_Proceedings_of_the_17th_NISTx2dNCSC_National_Computer_Security_Conference1994x22x29}\Autobibentry{Ravi Ganesan and Chris Davies. A new attack on random pronounceable password generators. In \textit{Proc. Proceedings of the 17th NIST{-}NCSC National Computer Security Conference}, 1994.}

\label{t:x28autobib_x22David_Ax2e_Huffman_and_othersA_method_for_the_construction_of_minimum_redundancy_codesProceedings_of_the_IRE_40x289x29x2c_ppx2e_1098x2dx2d11011952x22x29}\Autobibentry{David A. Huffman and others. A method for the construction of minimum redundancy codes. \textit{Proceedings of the IRE} 40(9), pp. 1098{--}1101, 1952.}

\label{t:x28autobib_x22Michael_Dx2e_Leonhard_and_VN_VenkatakrishnanA_comparative_study_of_three_random_password_generatorsIEEE_EITx2c_ppx2e_227x2dx2d2322007x22x29}\Autobibentry{Michael D. Leonhard and VN Venkatakrishnan. A comparative study of three random password generators. \textit{IEEE EIT}, pp. 227{--}232, 2007.}

\label{t:x28autobib_x22Randall_MonroePassword_Strengthx2c_XKCD_x23936on_the_webx3a_httpx3ax2fx2fwwwx2exkcdx2ecomx2f936x2f2011x22x29}\Autobibentry{Randall Monroe. Password Strength, XKCD \#936. on the web: http://www.xkcd.com/936/, 2011.}

\label{t:x28autobib_x22NISTAutomated_Password_GeneratorFederal_Information_Processing_Standards_Publication_Nox2e_1811993x22x29}\Autobibentry{NIST. Automated Password Generator. Federal Information Processing Standards Publication No. 181, 1993.}

\label{t:x28autobib_x22Claude_Ex2e_ShannonA_Mathematical_Theory_of_CommunicationBell_System_Technical_Journal_7x2c_ppx2e_379x2dx2d4231948x22x29}\Autobibentry{Claude E. Shannon. A Mathematical Theory of Communication. \textit{Bell System Technical Journal} 7, pp. 379{--}423, 1948.}

\label{t:x28autobib_x22Chris_StreetRIDYHEWx2e_The_RIDiculouslY_Huge_English_Wordliston_the_webx3a_httpx3ax2fx2fwwwx2ecodehappyx2enetx2fwordlistx2ehtmx22x29}\Autobibentry{Chris Street. RIDYHEW. The RIDiculouslY Huge English Wordlist. on the web: http://www.codehappy.net/wordlist.htm.}\end{SingleColumn}\end{AutoBibliography}

\postDoc
\end{document}